\documentclass{aastex61}

\usepackage{CJK}
\usepackage{amssymb}
\usepackage{multirow}
\usepackage{morefloats}
\usepackage{amsmath}
\usepackage{bigstrut}
\usepackage{booktabs}
\usepackage{natbib}
\usepackage{float}
\usepackage{graphicx,epsfig,fancyhdr,epsf,txfonts,epstopdf}
\usepackage{latexsym,bbm}
\usepackage{lineno}
\usepackage{color}
\usepackage{ulem}
\usepackage{threeparttable}
\usepackage{longtable}
\usepackage{hyperref}

\usepackage{lineno}

\received{***}
\revised{***}
\accepted{***}

\shortauthors{Li et al.}
\shorttitle{Modelling Periodic-Beaded Stripes}
\begin{document}
\large

\title{Coronal Diagnostics Via Modelling Periodic-Beaded Stripes of Solar Radio Bursts}

\correspondingauthor{Yao Chen}
\email{yaochen@sdu.edu.cn}

\author{Chuanyang Li}
\affiliation{Shandong Key Laboratory of Space Environment and Exploration Technology, College of Physics and Electronic Information, Dezhou University, Dezhou 253023, China}

\author{Yao Chen}
\affiliation{Center for Integrated Research on Space Science, Astronomy, and Physics, Institute of Frontier and Interdisciplinary Science, Shandong University, Qingdao 266237, China}
\affiliation{Shandong Key Laboratory of Space Environment and Exploration Technology, Institute of Space Sciences, Shandong University, Weihai 264209, China}

\author{Bing Wang}
\affiliation{Shandong Key Laboratory of Space Environment and Exploration Technology, Institute of Space Sciences, Shandong University, Weihai 264209, China}

\author{Yutong Li}
\affiliation{Shandong Key Laboratory of Space Environment and Exploration Technology, College of Physics and Electronic Information, Dezhou University, Dezhou 253023, China}

\author{Xiangliang Kong}
\affiliation{Center for Integrated Research on Space Science, Astronomy, and Physics, Institute of Frontier and Interdisciplinary Science, Shandong University, Qingdao 266237, China}
\affiliation{Shandong Key Laboratory of Space Environment and Exploration Technology, Institute of Space Sciences, Shandong University, Weihai 264209, China}

\author{Hao Ning}
\affiliation{Center for Integrated Research on Space Science, Astronomy, and Physics, Institute of Frontier and Interdisciplinary Science, Shandong University, Qingdao 266237, China}
\affiliation{Shandong Key Laboratory of Space Environment and Exploration Technology, Institute of Space Sciences, Shandong University, Weihai 264209, China}

\author{Sulan Ni}
\affiliation{School of Physics and Electronic Information, Yantai University, Yantai 264005, China}

\author{Shuwang Chang}
\affiliation{Center for Integrated Research on Space Science, Astronomy, and Physics, Institute of Frontier and Interdisciplinary Science, Shandong University, Qingdao 266237, China}

\author{Zichuan Li}
\affiliation{Shandong Key Laboratory of Space Environment and Exploration Technology, College of Physics and Electronic Information, Dezhou University, Dezhou 253023, China}

\author{Yang Gao}
\affiliation{Shandong Key Laboratory of Space Environment and Exploration Technology, College of Physics and Electronic Information, Dezhou University, Dezhou 253023, China}

\author{Zhe Cui}
\affiliation{Shandong Key Laboratory of Space Environment and Exploration Technology, College of Physics and Electronic Information, Dezhou University, Dezhou 253023, China}

\author{Li Deng}
\affiliation{National Space Science Center, Chinese Academy of Sciences, Beijing 100190, China}

\author{Jingye Yan}
\affiliation{National Space Science Center, Chinese Academy of Sciences, Beijing 100190, China}

\author{Fabao Yan}
\affiliation{Shandong Key Laboratory of Space Environment and Exploration Technology, Institute of Space Sciences, Shandong University, Weihai 264209, China}


\begin{abstract}
\large

Using high-resolution data from the Chashan Broadband Solar radio spectrometer at meter wavelengths (CBSm) of the Chinese Meridian Project-Phase II (CMP-II), \cite{Li2025} identified a novel fine spectral structure of solar radio bursts, termed periodic beaded stripes, and proposed a generation mechanism. Here we report additional events and develop a quantitative method to determine the physical conditions in the emission region. Periodic stripes tend to occur in the post-phase of flares and are associated with complex magnetic configurations. They repeat on sub-second timescales and show $\sim$0.1 s bead-like modulations, often accompanied by low-frequency absorptions. Modeling the chained stripes with linear kinetic theory of the double plasma resonance (DPR) instability constrains the source-region magnetic field to 0.2--1.7 G and the plasma density to (1--7) $\times 10^8$ cm $^{-3}$. The former follows the drift of individual stripes, and the latter tracks the overall trend. This study summarizes the key properties of periodic beaded stripes and establishes a quantitative DPR-based framework for coronal diagnostics.

\end{abstract}

\keywords{
\href{http://astrothesaurus.org/uat/1261}{Plasma astrophysics (1261)};
\href{http://astrothesaurus.org/uat/1490}{Solar electromagnetic emission(1490)};
\href{http://astrothesaurus.org/uat/1522}{Solar radio emission(1522)};
\href{http://astrothesaurus.org/uat/1339}{Radio bursts (1339)};
\href{http://astrothesaurus.org/uat/1544}{Space plasmas(1544)};
\href{http://astrothesaurus.org/uat/1993}{Solar coronal radio emission (1993)};
}

\section{Introduction}

Type IV solar radio bursts appear as broadband continua that occasionally contain diverse fine structures, including zebra patterns, intermediate drift bursts, spikes, fibers, and slowly drifting chains of narrowband fibers (e.g., \citealp{Kruger1984, Aurass1987, Mann1989, Chernov2006, Chernov2008, Feng2018}). These fine structures may arise from intrinsic emission mechanisms or from modulation by plasma inhomogeneities and low-frequency waves in the source region. Analysis of these structures probes microscopic flare dynamics, constrains eruption processes, and enables diagnostics of coronal conditions.

Recently, \cite{Li2025} analyzed high-resolution observations from the Chashan Broadband Solar radio spectrometer at meter wavelengths (CBSm: \citealp{Chang2024}), an instrument of the Chinese Meridian Project-Phase II (CMP-II), and identified a novel fine spectral structure in type-IV solar radio bursts, termed periodic beaded stripes. These structures consist of periodic narrow-band stripes with recurrence times shorter than 1 s. Each stripe ends as the next starts, forming an end-to-end sequence, and often drifts from high to low frequencies, accompanied by absorption features. In some cases, the stripes display periodic beaded (pearl-like) enhancements with a characteristic timescale of $\sim$0.1 s. \cite{Li2025} proposed that the periodic stripes arise from successive excitation of upper-hybrid (UH) harmonics driven by the double plasma resonance (DPR) instability, and attributed the beaded substructures to modulation by low-frequency magnetohydrodynamic (MHD) waves.

The DPR- plasma emission process serves as a well-accepted framework for interpreting type-IV bursts and zebra patterns. It involves trapped electrons which, in the regime of $\omega_{pe} \gg \Omega_{ce}$, drive the DPR instability and excite electrostatic UH waves. These waves are strongly amplified when the resonance condition is satisfied, i.e., when the UH frequency approaches an integer multiple of the electron cyclotron frequency ($\omega_{UH} \approx s\Omega_{ce}$). In an inhomogeneous plasma, variations in density or magnetic field strength modify the local ratio $\omega_{pe}/\Omega_{ce}$ and shift the resonance condition, thereby modulating the UH wave growth rate. Mode conversion then transforms the amplified UH waves into escaping fundamental or harmonic emission (e.g., \citealp{Zheleznyakov1975, Zlotnik2013, Li2019, Li2021, Ni2020}).

This paper analyzes recent CBSm observations of periodic beaded stripes and summarizes the observational properties. Linear kinetic theory quantifies the growth rates of UH waves driven by the DPR instability and reproduces the observed chain-like and bead-like structures. The results establish a quantitative diagnostic framework for coronal magnetic and plasma parameters. Section \ref{sec:2} describes the observations and the proposed generation mechanism. Section \ref{sec:3} presents the modeling and diagnostic results. The final section is summary.

\section{Periodic Beaded stripes and the Generation Mechanism}
\label{sec:2}
The spectral data were obtained from CBSm in the range of 90--600 MHz. The CBSm system is located at the Chashan Solar Observatory (CSO), which is managed by the Institute of Space Sciences of Shandong University. It works with a 12 m parabolic reflector, a dual linear-polarized log-periodic feed, and a high-precision tracking platform. The default resolution is 76.29 kHz in frequency and 0.84 ms (up to 0.21 ms) in time. The dynamic range of the system is $\sim$60--65 dB and its sensitivity is $\sim$1 SFU for an integration time of 1 ms and a bandwidth of 100 kHz (\citealp{Chang2024}). It has detected hundreds of meter-wave bursts, including 85 type II bursts till January 26, 2026, along with numerous bursts exhibiting fine structures (\href{http://47.104.87.104/MWRS/RadioBurstEvent/typeII/typeIIburst_show.html}
{gallery of CSO type II bursts}). Data from CBSm have been used to study various types of solar radio bursts, such as type I and II bursts, quasi-periodic pulsations, etc. (e.g., \citealp{Hou2023, Shi2024, Yang2025, LiDong2025, Cui2025}).

Additional data came from DAocheng Radio Telescope (DART: \citealp{Yan2023}) and the Solar Dynamics Observatory (SDO: \citealp{Pesnell2012}). DART performs aperture-synthesis imaging at 16 discrete frequencies between 150 and 450 MHz (149, 164, 190, 205, 223, 238, 285, 300, 309, 324, 366, 381, 399, 414, 432, and 447 MHz) and provides an angular resolution ranging from 5{$^{\prime}$ (149 MHz) to 1.5{$^{\prime}$ (450 MHz) (e.g., \citealp{Hou2025, Li2025, Wang2025}). Magnetograms from the Helioseismic and Magnetic Imager (HMI: cadence 45 s, pixel size 0.5{$^{\prime\prime}$; \citealp{Schou2012}) and multi-wavelength extreme-ultraviolet (EUV) images from the Atmospheric Imaging Assembly (AIA: cadence 12 s, pixel size 0.6{$^{\prime\prime}$; \citealp{Lemen2012}), both onboard SDO, were used to characterize the magnetic and coronal environment of the associated active region (AR).

The next section presents periodic stripes observed by CBSm since 2022, including events that display clear beaded substructures, and outlines the proposed generation mechanism. The selected events include those on 2024 May 5, 8, and 9, and 2025 May 14 and October 13. Some chained stripes from the 2024 May 5 and 8 events were previously reported by \cite{Li2025}.

\subsection{Characteristics of Periodic Beaded Stripes }

Figure \ref{Fig:figure1} shows the event on 2024 May 5, in which abundant narrow-band stripes appear at the low-frequency end of the type-IV continuum. The GOES soft X-ray flux indicates that the event occurred in the pre-flare stage of an M8.4-class flare peaking at 01:27:00 UT. The zoomed-in spectra (Figures \ref{Fig:figure1}(b) and (c)) display diverse morphologies. The stripes appear either as isolated or as chains, sometimes overlapping. The chained stripes tend to drift from high to low frequencies and show absorption on the low-frequency side. Each stripe starts as the preceding one ends. Figures \ref{Fig:figure1}(d)--(i) present six typical chained stripes. The stripes exhibit periods of 0.5--1.5 s, bandwidths of 0.5--1.5 MHz, durations of 0.5--1.5 s, and frequency drift rates of --0.7 to --2 MHz/s.

The event on 2024 May 8 includes three episodes of enhanced emission (Figure \ref{Fig:figure2}). The first two correspond to an X1.0 flare (peaking at 21:40:00 UT) and an M9.9 flare (peaking at 22:27:00 UT). The third occurs in the post-phase of the second flare and shows weaker intensity, a narrower bandwidth (250--500 MHz), and abundant narrow-band stripes (Figures \ref{Fig:figure2}(b) and (c)). Figures \ref{Fig:figure2}(d)--(i) reveal periodic intensity enhancements along some stripes, resulting in a distinct beaded pattern. Each stripe is accompanied by absorption on the low-frequency side, and the absorption bandwidth typically exceeds that of the emission.

We measured the main parameters of each chained stripes. The stripes within a chain have recurrence periods of 0.4--1 s, bandwidths of 2--5 MHz, durations of 0.4--2 s, and frequency drift rates of --2 to --13 MHz/s. The beaded substructures show frequency separations of $\sim$0.8 MHz, bandwidths of $\sim$0.6 MHz, and temporal spacings of $\sim$0.1 s.

Similar narrow-band chained stripes and beaded structures were observed in the 2024 May 9 event as well (Figure \ref{Fig:figure3}). Within these chains, the stripes have periods of 0.2--1 s, bandwidths of 1--5 MHz, durations of 0.4--1 s, and frequency drift rates of --6 to --10 MHz/s. The beaded substructures show frequency separations below 1 MHz and temporal spacings of $\sim$0.1 s.

The 2025 May 14 event occurred in the post-phase of an X2.7 flare peaking at 08:25:00 UT (Figure \ref{Fig:figure4}). The stripes show periods of 0.5--1 s, bandwidths of 0.5--2.5 MHz, durations of 0.5--2 s, and drift rates of --1.2 to --2 MHz/s. The 2025 October 13 event arose during the post-phase of an M1.9 flare peaking at 05:26:00 UT (Figure \ref{Fig:figure5}). The stripes exhibit periods of 0.5--1 s, bandwidths of 0.5--2.5 MHz, durations of 0.3--1.2 s, and drift rates of --3.7 to 0.7 MHz/s. Figure \ref{Fig:figure5}(e) shows a rare case in which both individual stripes and the chains drift toward higher frequencies.

Here we summarize the main properties of periodic beaded stripes: (1) individual stripes often drift from high to low frequencies, whereas the chains show variable drift directions; adjacent stripes form an end-to-end sequence and are accompanied by absorptions on the low-frequency side; (2) stripes exhibit sub-second periodicity ($<$1 s), and the beaded structures show periods of $\sim$0.1 s; (3) these events arise mainly during post-phases of flares and only occasionally during pre-phases. In addition, DART radio imaging and SDO observations further indicate that the associated ARs possess complex magnetic topologies (Figures \ref{Fig:figure6} and \ref{Fig:figure7}). These properties place strong constraints on models of periodic beaded stripes.

The radio sources co-align with the AR nearby, although some displacement exists (see Figure \ref{Fig:figure6}). It is mainly due to the projection effect since the sources are located high above the solar surface, at an altitude of $\sim$0.2--0.3 $R_{\odot}$. This can be told from the limb source (see Figure \ref{Fig:figure6} (d)). Other effects such as scattering, diffraction, and reflection also contribute.

\subsection{Generation Mechanism of Periodic Beaded Stripes}

Taking the 2024 May 9 event as a representative case, this section examines the associated AR and outlines the proposed formation mechanism.

Figures \ref{Fig:figure7}(a)--(c) present SDO observations of the associated AR near 00:00 UT on May 9, with co-spatial DART radio imaging overlaid. The radio source has a brightness temperature exceeding 10$^9$ K and located southeast of AR 13664. AIA images reveal brightening loops beneath the source, while HMI line-of-sight magnetograms show a complex magnetic topology with significant shear motions. Using the nonlinear force-free field (NLFFF) extrapolation method, we reconstruct the coronal magnetic field and identify highly twisted structures within the AR associated with the sources (Figure \ref{Fig:figure7}(d)). According to the extrapolation, the field strength around decreases rapidly with height, reaching a level of few Gauss at the possible altitude of the sources (180--250 Mm, $\sim$0.3 $R_{\odot}$). Weaker field may exist if the sources originate from the null- or X-type region of the magnetic field. The plasma density at this height of solar ARs may be at a level of 10$^8$ cm$^{-3}$ (see, e.g., \citealp{Aschwanden2004, Aschwanden2019}). The combination of these parameters yield  $\omega_{pe}/\Omega_{ce}$ to be several tens.


Following \cite{Li2025}, we interpret the chained stripes within the DPR framework. The flare occurs in the AR with a multipolar configuration, above which overlying loops exist. Such configuration may contain the X-type region where magnetic field is weak. This forms an effective magnetic trap, which can trap radio-emitting energetic electrons. In overdense plasmas with $\omega_{pe}/\Omega_{ce} \gg 1$, the trapped electrons drive the DPR instability and excite UH waves at frequency close to $\omega_{UH}$ when $\omega_{UH} \approx s \Omega_{ce}$, where $s$ is a positive integer. Temporal variations in plasma density ($n_0$) and/or magnetic field strength ($B_0$) modulate the values of $\omega_{pe}/\Omega_{ce}$. As shown in the cartoons of Figures \ref{Fig:figure7}(e) and (f), a unit change in $\omega_{pe}/\Omega_{ce}$ shifts the harmonic number from $s$ to $s+1$ or $s-1$ and generates successive UH harmonics at ($s \pm 1$) $\Omega_{ce}$. Nonlinear wave--wave coalescence then converts UH waves into electromagnetic emission.

The absorption feature results from negative growth of UH waves. Wave dispersion analysis (e.g., \citealp{Li2019}) shows that a small frequency shift can switch efficient wave amplification to strong wave damping, and vice versa, naturally producing radiation-absorption pairs. The periodic bead-like structures within the stripes arise from modulation of the UH waves by low-frequency MHD waves, such as Alfv\'{e}n waves or fast/slow magnetosonic waves.

The following section evaluates the DPR-driven growth rate of UH waves and applies the model to the observations to constrain coronal parameters.

\section{Modeling and Coronal Diagnostics}
\label{sec:3}
The plasma density $n_0$ is set to decrease from $5.6 \times 10^8$ to $5.4 \times$ $10^8$ cm $^{-3}$, and the magnetic field strength $B_0$ from 0.95 to 0.83 G (Figure \ref{Fig:figure8}(a)), as $\omega_{pe}/\Omega_{ce}$ increases from 80 to 90 over 5 s (Figure \ref{Fig:figure8}(b)).

The model assumes a uniformly magnetized plasma. Background electrons follow a Maxwellian distribution ($f_0$), and energetic electrons with a loss-cone distribution ($f_e$), forming the particle distribution in ($u_{\parallel}, u_{\perp}$) space shown in Figure \ref{Fig:figure8}(c). The background electron temperature $T_0 = 2$ MK, the mean energetic electron speed $v_e$=0.3 c, and the energetic-to-background electron density ratio is $n_e/n_0 = 1\%$. The Appendix details the distribution functions, dispersion relation, and the growth rate ($\gamma$) of Z-mode, which becomes the UH mode in the perpendicular propagation limit.

Figure \ref{Fig:figure8}(d) shows the Z-mode dispersion relation at $\omega_{pe}/\Omega_{ce}$ = 80 and $\theta$ = 90$^\circ$ (red circles). It agrees closely with the kinetic Bernstein dispersion relation (black plus signs) and the perpendicular UH mode dispersion relation (blue curve). Figure \ref{Fig:figure8}(e) presents the Z-mode growth rate in ($\omega, \theta$) space for $\omega_{pe}/\Omega_{ce}$ = 80, with dominant growth concentrated near perpendicular propagation, consistent with the UH limit. The maximum growth rate reaches 2.1 $\Omega_{ce}n_e/n_0$ at ($84.9 \Omega_{ce}$, $89.5^\circ$). Figure \ref{Fig:figure8}(f) shows the frequency dependence of $\gamma$ at $\theta = 89.5^\circ$.

\subsection{DPR Instability: Model for Periodic Beaded Stripes }
\label{sec:3.1}

Figures \ref{Fig:figure9}(a) and (b) show the temporal evolution of the UH mode maximum growth rate ($\gamma_{max}$) and its corresponding frequency ($\omega_{max}$) and propagation angle ($\theta_{max}$) as functions of $\omega_{pe}/\Omega_{ce}$. The $\gamma_{max}$ remains nearly constant with weak quasi-periodic modulation, while the $\omega_{max}$ jumps between successive integer harmonics that satisfy $\omega_{UH} \approx s\Omega_{ce}$ with $s$ = 85--96. Assuming nonlinear wave--wave coalescence ($\rm UH + UH \rightarrow H$), the resulting spectrum in Figure \ref{Fig:figure9}(c) exhibits a stripe bandwidth of $\sim$5 MHz and a recurrence time of $\sim$0.5 s, with stripes connecting continuously in time, consistent with the observed chained stripes.

Varying the temporal profiles of $n_0$ and $B_0$ generates chained stripes with nearly constant frequencies (Figures \ref{Fig:figure9}(d)--(f)) or upward frequency drift (Figures \ref{Fig:figure9}(g)--(i)), reproducing the diverse observed drift patterns.

Figure \ref{Fig:figure10} presents the radiation spectrum derived from the UH mode growth rate for $\omega_{pe}/\Omega_{ce}$ varying from 85 to 87 ($\approx$1 s). It shows growth concentrated near discrete frequencies 91, 92, and 93 $\Omega_{ce}$ (panel (a2)), consistent with Figure \ref{Fig:figure9}(b), and the resulting spectrum after mode conversion (panel (a3)) organizes the UH harmonics into periodic chained stripes.

The growth rate scales as $\gamma \propto \Omega_{ce} n_e/n_0$, so changes in $B_0$ or $n_e$ directly modulate wave growth and radiation intensity. Figures \ref{Fig:figure10}(b1)--(b3) shows that a 10 Hz sinusoidal perturbation in $n_e$ with 20\% amplitude produces a beaded pattern along the stripes. Figures \ref{Fig:figure10}(c1)--(c3) shows that small changes in the magnetic-field direction by 0.05$^\circ$ around $\theta_{max}$ modify the growth rate and shift the emission frequency, producing wavy stripes similar to those occasionally observed (e.g., 2024 May 5, 01:12:45--01:13:00 UT, Figure \ref{Fig:figure1}(g)).

\subsection{Periodic Beaded Stripes as a Diagnostic of Coronal Parameters}

The model enables direct diagnostics of the source-region magnetic field and plasma density. Stripe frequencies constrain the electron density, while the spacing between adjacent stripes determines $\Omega_{ce}$ and, consequently, $B_0$.

For the 2024 May 9 event (Figure \ref{Fig:figure11}(a)), the frequency difference between the leading and trailing edges of consecutive stripes ranges from 1.9 to 3.4 MHz, which yields magnetic fields of 0.7--1.2 G without clear trend. According to our previous PIC simulations of DPR-driven type IV bursts, the harmonic emission is stronger than the fundamental emission by $\sim$2 orders of magnitude (\citealp{Li2021}), so we first assume the stripes to be second-harmonic with frequency being $\sim$2$\omega_{UH}$, we have the electron density varies from 5.7 to 5.9 $\times \ 10^8$ cm$^{-3}$ (Figure \ref{Fig:figure11}(d)) along the radio chain. If assuming the radio bursts to be fundamental, we get a density of (2.28--2.38) $\times \ 10^9$ cm$^{-3}$. From $\omega_{UH} \approx s\Omega_{ce}$, the excited UH waves span harmonic numbers $s$ = 60--120 (Figure \ref{Fig:figure11}(c)), with small relative changes in $\Omega_{ce}$ producing amplified shifts in $s$, since $\omega_{UH} \gg s\Omega_{ce}$.

The same procedure is applied to all events, with six chained stripes selected per event (Figures \ref{Fig:figure1}--\ref{Fig:figure5}(d)--(i)). Assuming the stripes correspond to second-harmonic emission, Figures \ref{Fig:figure12}(a1)--(e2) show results for individual events, and Figures \ref{Fig:figure12}(f1) and (f2) summarizes all thirty chains. Magnetic fields are weak, ranging from 0.2 to 1.7 G, with no clear trend, suggesting origin in intersecting multipolar loops above sunspots (Figures \ref{Fig:figure7}(c) and (d)). Electron densities span (1--7) $\times$ $10^8$ cm$^{-3}$ and follow the overall trend.


\section{Summary}
\label{sec:4}
Using high temporal and frequency-resolution observations from the CBSm system of CMP-II, multiple periodic narrow-band stripes in type IV continua were analyzed, several showing clear beaded substructures. These events occur mainly during post-phase of flares and occasionally before flares, but not at peak. The stripes drift from high to low frequencies with periods of 0.2--1.5 s and connect end-to-end in time, while individual stripes last 0.3--2 s and span 0.3--5 MHz. Beaded stripes show additional modulation on $\sim$0.1 s timescales, spacing below 1 MHz, and accompanied by absorptions on the low-frequency side. DART imaging combined with SDO observations links the radio sources to complex magnetic structures in active regions.

A diagnostic framework based on DPR theory reproduces the global drift, periodic structure, and end-to-end continuity of stripes by prescribing density and magnetic field evolution profiles, while perturbations in density or magnetic field produce beaded and distorted morphologies. Applying this model to thirty stripe-chains over five days yields source-region magnetic fields of 0.2--1.7 G and plasma densities of (1--7) $\times$ $10^8$ cm$^{-3}$. These results support the model proposed by \cite{Li2025}, indicate that periodic stripes arise from multi-harmonic UH waves with low-frequency MHD waves modulating the beaded substructures, and provide a novel method for diagnosing coronal magnetic field and plasma density.

\section*{Acknowledgements}
This study is supported by the Scientific Research Fund of Dezhou University (4022504002, 4022504003, 3012304024), the National Natural Science Foundation of China (NNSFC) grants (Nos. 12103029, 12203031, 12233005), the Natural Science Foundation of Shandong Province (NSFSP) grants (Nos. ZR2021QA079, ZR2023QA141, ZR2025QC1503, ZR2024QA212), the National Key R\&D Program of China under grant 2022YFF0503002 (2022YFF0503000). We acknowledge the use of data from the Chinese Meridian Project. We thank the teams of CSO, DART, GOES, and SDO for making their data available to us. The authors are grateful to the anonymous referee for the valuable comments.

\newpage
\appendix

\section{The Growth Rate and Dispersion Tensor of Z-mode Instability for Warm Plasmas}

This study adopts the general kinetic dispersion relation for small-amplitude waves propagating in a uniform, magnetized, warm plasma (see, e.g., \citealp{Baldwin1969, Wu1985}), which represents the linear solution of the collisionless Vlasov-Maxwell system. The plasma consists of two electron components: a background warm Maxwellian distribution $f_0$ with number density $n_0$ and an energetic loss-cone distribution $f_e$ with density $n_e \ll n_0$, so thermal electrons determine the wave modes, whereas energetic electrons drive the instability. The distributions are given by
\begin{equation}
 f(u_{\perp},u_{\parallel})=\frac{n_e}{n_0}f_e+\left(1-\frac{n_e}{n_0}\right)f_0,
\end{equation}
\begin{equation}
 f_0=\frac{1}{(2\pi)^{3/2}v_0^{3}}\exp\left(-\frac{u^{2}}{2v_0^{2}}\right),
\end{equation}
\begin{equation}
 f_e=\frac{1}{(2\pi)^{3/2}v_e^{3}A}\exp\left(-\frac{u^{2}}{2v_e^{2}}\right)\left(1- \rm{tanh}\frac{\mu -\mu_0 }{\delta}\right),
\end{equation}
\begin{equation}
 A=1 + \frac{\delta}{2}\left(\rm{ln}(\rm{cosh}\frac{1+\mu_0 }{\delta} - \rm{ln}(\rm{cosh}\frac{1-\mu_0 }{\delta})\right),
\end{equation}
where $f$ denotes the total electron distribution function, $v_e$ is the mean velocity of energetic electrons, $u = p/mc$ is the normalized momentum, and $\mu = p_{\parallel}/p_{\perp}$ is the pitch-angle cosine. The loss-cone boundary $\mu_0$ = 0.6 (corresponding to a loss-cone angle of 53${^\circ}$), and $\delta$ = 0.25 controls smoothing at the loss-cone edge.

The wave frequency takes the form $\omega = \omega_r + i\gamma$, where $\gamma$ denotes the growth rate, $\omega_r$ is determined by the dispersion relation using the following fluid approximation of warm plasmas for Z-mode,
\begin{equation}
\text{Re}\overleftrightarrow{\Lambda}(\vec{k},\omega_r)=
\left( \begin{array}{ccc}
 -N^2\cos^2\theta & 0 & N^2\sin\cos\theta \\
 0 & -N^2 & 0 \\
 N^2\sin\cos\theta & 0 & -N^2\sin^2\theta
\end{array}
\right )+\overleftrightarrow{\epsilon}=0,
\end{equation}
\begin{equation}
\overleftrightarrow{\epsilon}=\overleftrightarrow{I}-\frac{\omega_{pe}^2}{\omega_r^2}\frac{\overleftrightarrow{C}_e}{\Delta_e}, \Delta_e=(1-\frac{\Omega_{ce}^2}{\omega_r})(1-3N^2v_0^2\cos^2\theta)-3N^2v_0^2\sin^2\theta,
\end{equation}
\begin{equation}
\overleftrightarrow{C}_e=
\left( \begin{array}{ccc}
 1-3N^2v_0^2\cos^2\theta & -i\frac{\Omega_{ce}}{\omega_r}(1-3N^2v_0^2\cos^2\theta) & 3N^2v_0^2\sin\theta\cos\theta \\
 i\frac{\Omega_{ce}}{\omega_r}(1-3N^2v_0^2\cos^2\theta) & 1-3N^2v_0^2 & i\frac{\Omega_{ce}}{\omega_r}3N^2v_0^2\sin\theta\cos\theta \\
 3N^2v_0^2\sin\theta\cos\theta & -i\frac{\Omega_{ce}}{\omega_r}3N^2v_0^2\sin\theta\cos\theta & 1-\frac{\Omega_{ce}^2}{\omega_r^2}-3N^2v_0^2\sin^2\theta
\end{array}
\right ),
\end{equation}
 where $N=kc/\omega_r$ and $k$ are the refractive index and the wave number, respectively, and $\theta$ is the angle of propagation (i.e., the angle between $\vec{k}$ and $\vec{B}$), and $v_0=\sqrt{k_BT_0/m_e}$ is the thermal velocity of background electrons.

Under the assumption of $\omega_r \gg \left|\gamma\right|$, the growth rate $\gamma$ is given by
\begin{equation}
\gamma=-\frac{\text{Im}\Lambda(\vec{k},\omega_r)}{\frac{\partial}{\partial\omega_r}\text{Re}\Lambda(\vec{k},\omega_r)}.
\end{equation}

$\text{Im}\Lambda(\vec{k},\omega_r)$ is the imaginary part of the kinetic dispersion relation (e.g., \citealp{Baldwin1969, Li2019}), denotes
\begin{equation}
\begin{split}
\text{Im}\Lambda(\vec{k},\omega_r)=&2\pi\frac{\omega_{pe}^2}{\omega_r^2}\int_{-\infty}^{+\infty}du_{\parallel}\int_{0}^{+\infty}du_{\perp}\Bigg\{\frac{u_{\parallel}}{\gamma_L}\left(u_\perp\frac{\partial}{\partial u_\parallel}-u_\parallel\frac{\partial}{\partial u_\perp}\right)\times f(u_\perp,u_\parallel)\hat{e}_z\hat{e}_z+\\
&\omega_r\left[\frac{\partial}{\partial u_\perp}+\frac{k_\parallel}{\gamma_L\omega_r}\left(u_\perp\frac{\partial}{\partial u_\parallel}-u_\parallel\frac{\partial}{\partial u_\perp}\right)\right]\times f(u_\perp,u_\parallel)\sum_{n=-\infty}^{\infty}\frac{\overleftrightarrow{T_n}(b)}{\gamma_L\omega_r-n\Omega_{ce}-k_\parallel u_\parallel}\Bigg\}
\end{split},
\end{equation}
\begin{equation}
\overleftrightarrow{T_n}(b)=
\left( \begin{array}{ccc}
\frac{n^2\Omega_{ce}^2}{k_\perp^2}J_n^2(b) & -i\frac{n\Omega_{ce}}{k_\perp}u_\perp J_n(b)J_n^\prime(b) & \frac{n\Omega_{ce}}{k_\perp}u_\parallel J_n^2(b)\\
i\frac{n\Omega_{ce}}{k_\perp}u_\perp J_n(b)J_n^\prime(b) & u_\perp^2J_n^{\prime2}(b) & iu_\perp u_\parallel J_n(b)J_n^\prime(b)\\
\frac{n\Omega_{ce}}{k_\perp}u_\parallel J_n^2(b) & -iu_\perp u_\parallel J_n(b)J_n^\prime(b) & u_\parallel^2J_n^2(b)
\end{array}
\right ),
\end{equation}
where $u_\parallel=p_\parallel /m_e=\gamma_L v_\parallel$, $u_\perp=p_\perp /m_e=\gamma_L v_\perp$ denote the parallel and perpendicular momenta normalized by $m_e$, $\omega_{pe}=\sqrt{n_ee^2 /m_e\varepsilon_0}$ is the plasma frequency, $\gamma_L=\left(1-\frac{v^2}{c^2}\right)^{-1/2}$ is the Lorentz factor, $f(u_\perp,u_\parallel)$ is the total distribution function of electrons, and $J_n(b)$ the first-kind Bessel function of the \emph{n}th order, $J_n^\prime(b)$ is its partial derivative with respect to $b \ (=k_\perp u_\perp /\Omega_{ce})$.


 \begin{figure*}
   \centerline{\includegraphics[width=0.9\textwidth]{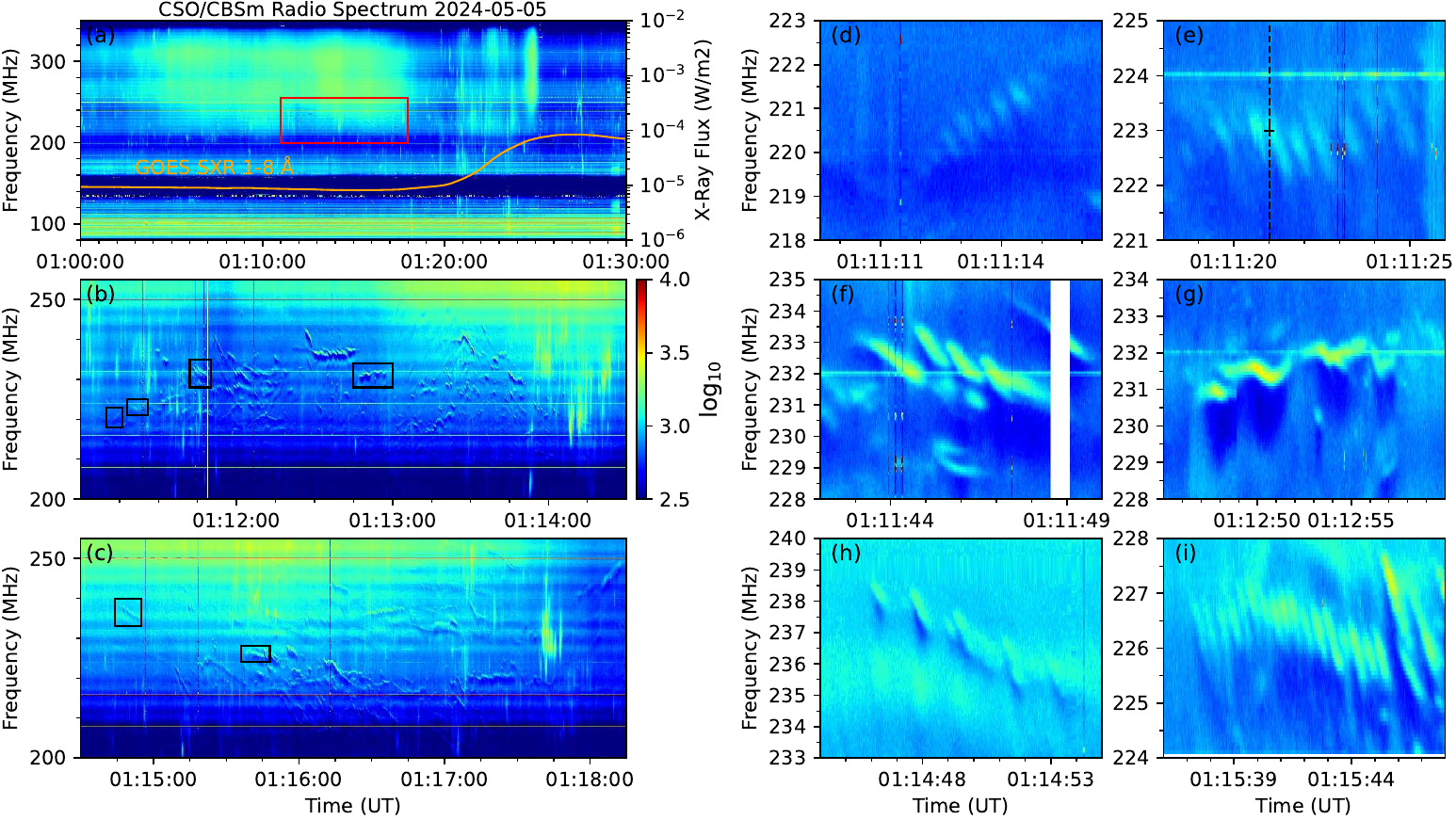}
              }
              \caption{Solar radio dynamic spectrum of the 2024 May 5 event observed by CBSm. (a) Overview of the spectrum (80--350 MHz, 01:00--01:30 UT) with overlaid GOES 1--8 \AA \ soft X-ray flux (yellow curve). (b) and (c) Zoomed-in views of the selected region in panel (a) (red box). (d)--(i) Zoomed-in views of the selected regions in panels (b) and (c) (black boxes). The black dashed line and plus sign in panel (e) mark the DART imaging time and frequency shown in Figure \ref{Fig:figure6}(a).
              }
   \label{Fig:figure1}
   \end{figure*}

 \begin{figure*}
   \centerline{\includegraphics[width=0.9\textwidth]{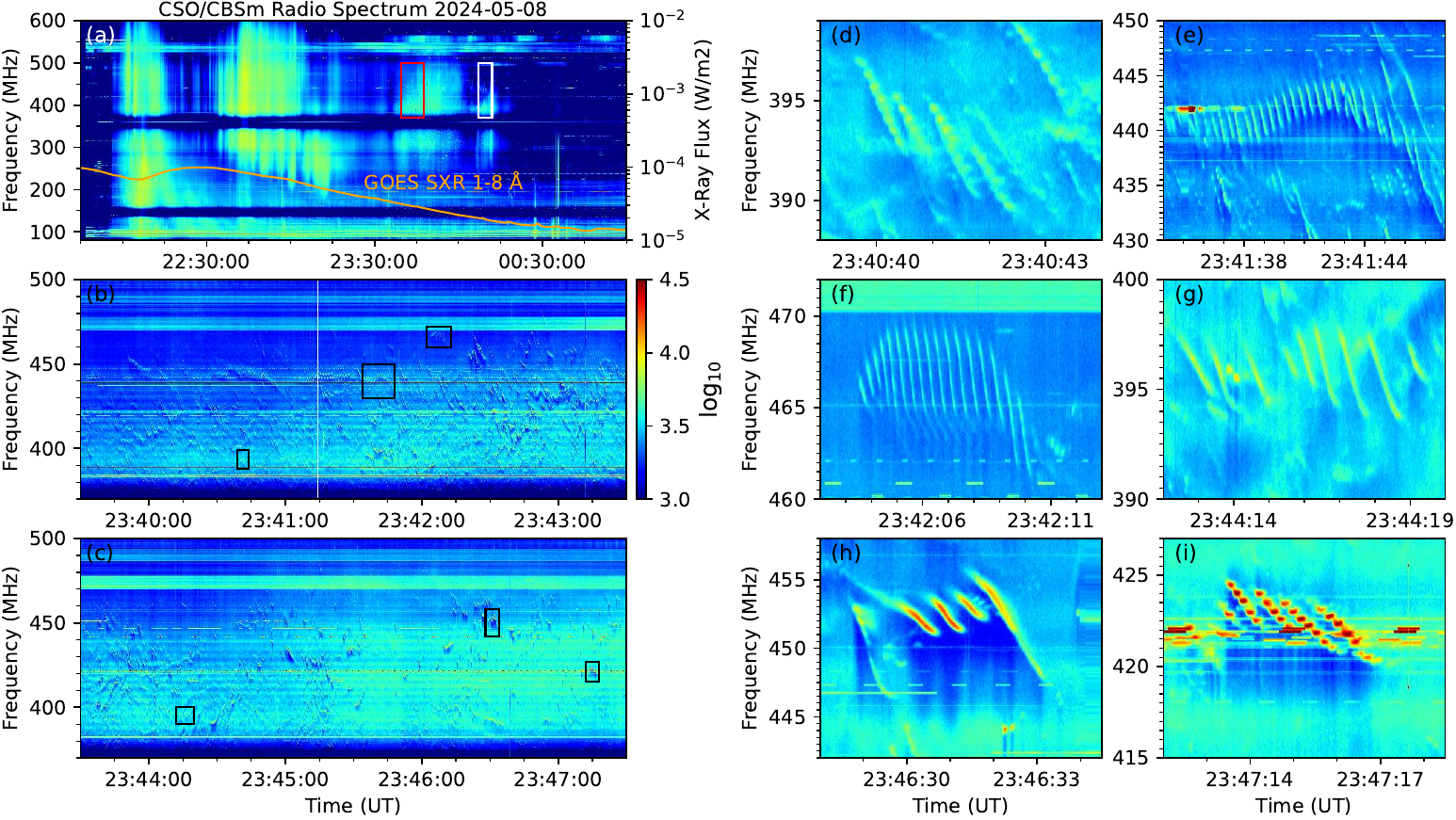}
              }
              \caption{Same as Figure \ref{Fig:figure1}, but for the 2024 May 8 event. Zoomed-in view of the white box is shown in Figure \ref{Fig:figure3}.
              }
   \label{Fig:figure2}
   \end{figure*}

 \begin{figure*}
   \centerline{\includegraphics[width=0.7\textwidth]{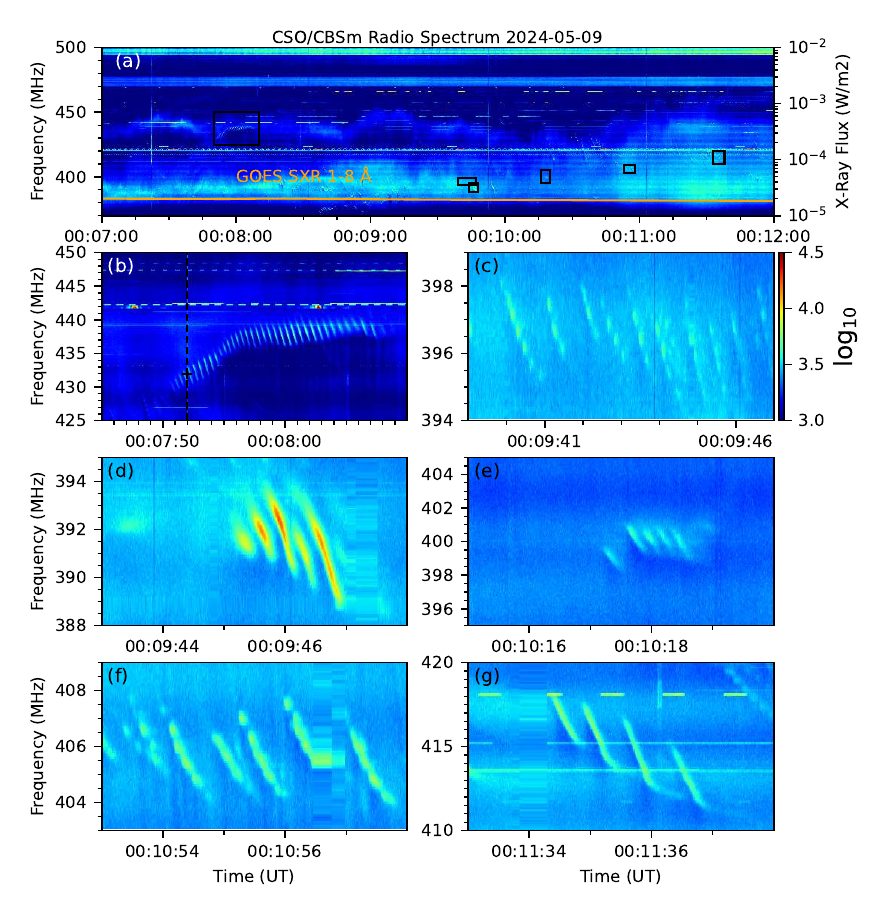}
              }
              \caption{Solar radio dynamic spectrum of the 2024 May 9 event observed by CBSm. (a) Zoomed-in view of the selected region in Figure \ref{Fig:figure2}(a) (white box), covering 370--500 MHz during 00:07--00:12 UT with overlaid GOES 1--8 \AA \ soft X-ray flux (yellow curve). (b)--(g) Zoomed-in views of the selected region in panel (a) (black boxes). The black dashed line and plus sign in panel (b) mark the DART imaging time and frequency shown in Figure \ref{Fig:figure7}.
              }
   \label{Fig:figure3}
   \end{figure*}

 \begin{figure*}
   \centerline{\includegraphics[width=0.9\textwidth]{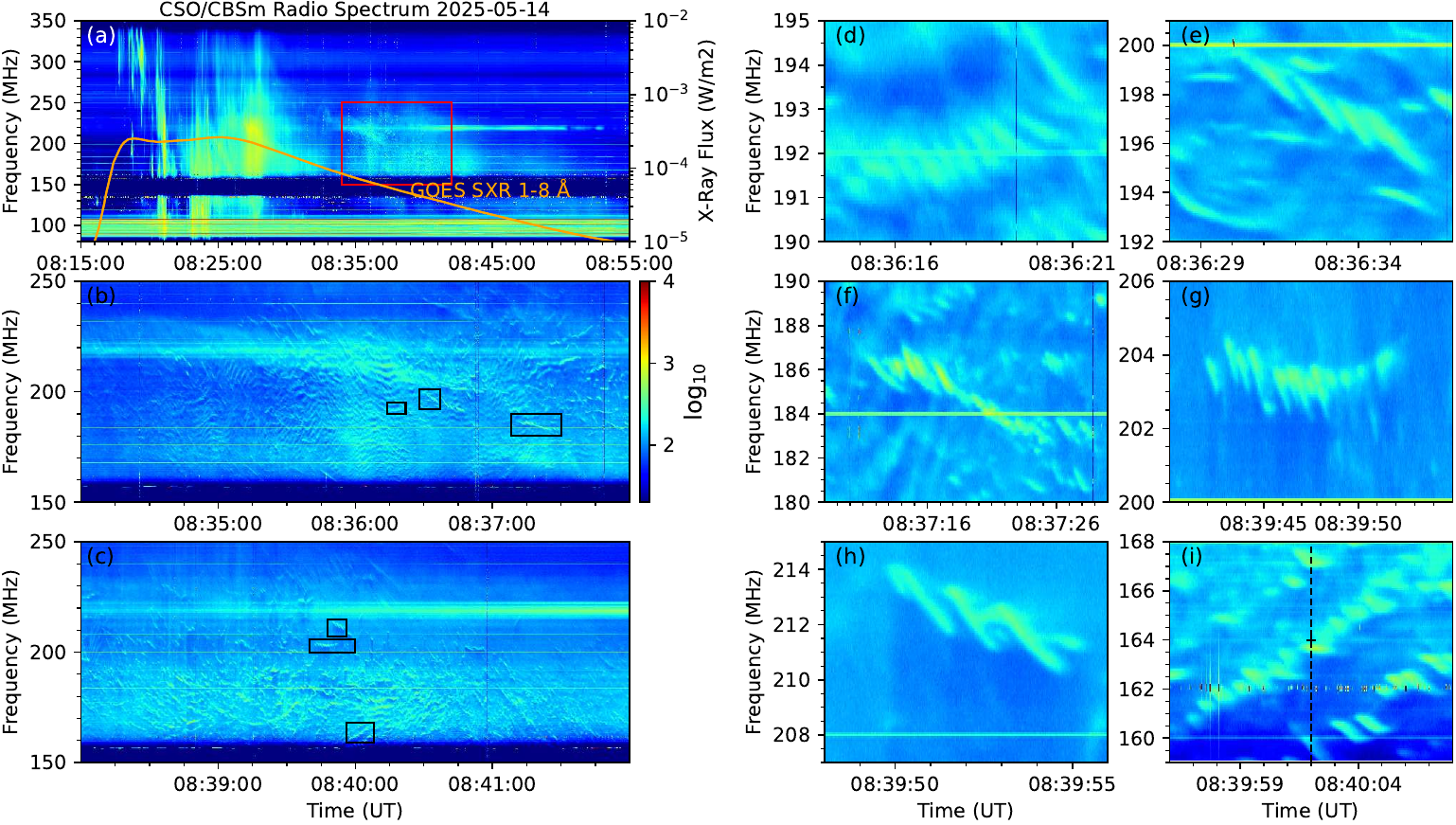}
              }
              \caption{Same as Figure \ref{Fig:figure1}, but for the 2025 May 14 event. The black dashed line and plus sign in panel (i) mark the DART imaging time and frequency shown in Figure \ref{Fig:figure6}(d).
              }
   \label{Fig:figure4}
   \end{figure*}

 \begin{figure*}
   \centerline{\includegraphics[width=0.9\textwidth]{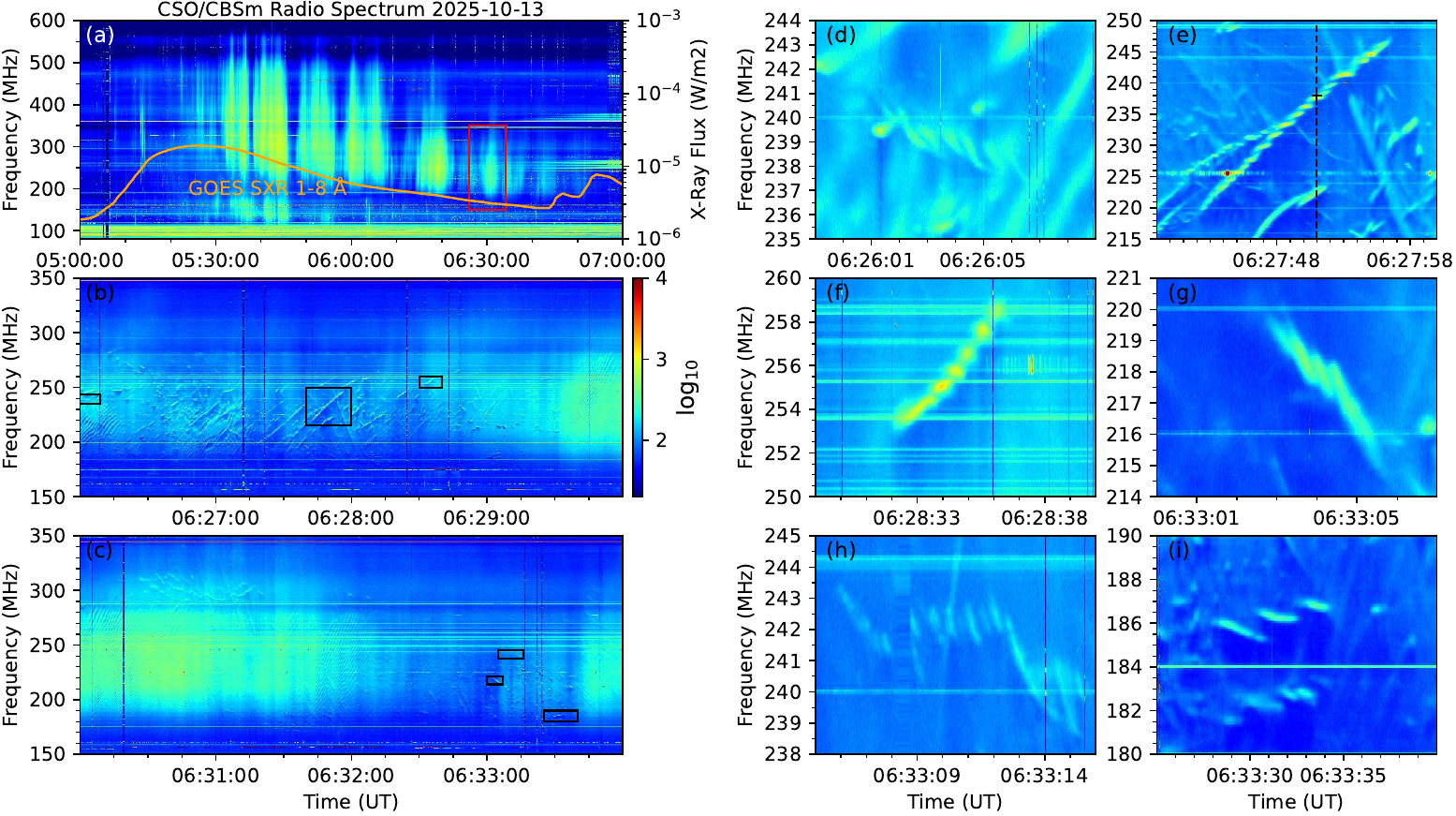}
              }
              \caption{ Same as Figure \ref{Fig:figure1}, but for the 2025 October 13 event. The black dashed line and plus sign in panel (e) mark the DART imaging time and frequency shown in Figure \ref{Fig:figure6}(g).
              }
   \label{Fig:figure5}
   \end{figure*}

 \begin{figure*}
   \centerline{\includegraphics[width=0.9\textwidth]{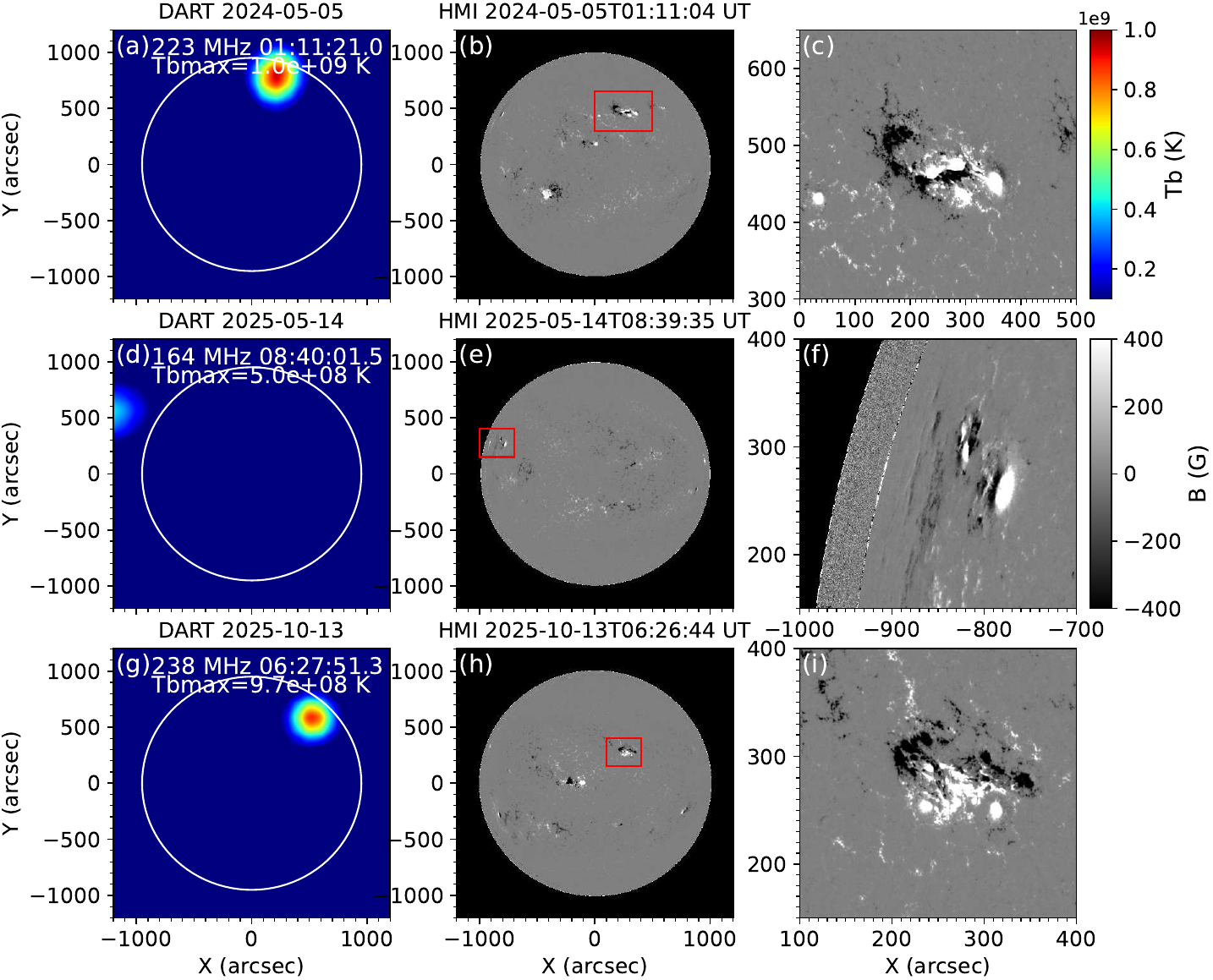}
              }
              \caption{DART radio images and simultaneous HMI line-of-sight magnetograms of the ARs associated with narrow-band chained stripes on 2024 May 5, 2025 May 14, and 2025 October 13. The top, middle, and bottom rows correspond to the three events, respectively. The right column shows zoomed-in views of the red boxes. The DART imaging times and frequencies for these three events are marked in the dynamic spectra of Figures \ref{Fig:figure1}, \ref{Fig:figure4}, and \ref{Fig:figure5}.
              }
   \label{Fig:figure6}
   \end{figure*}

 \begin{figure*}
   \centerline{\includegraphics[width=0.9\textwidth]{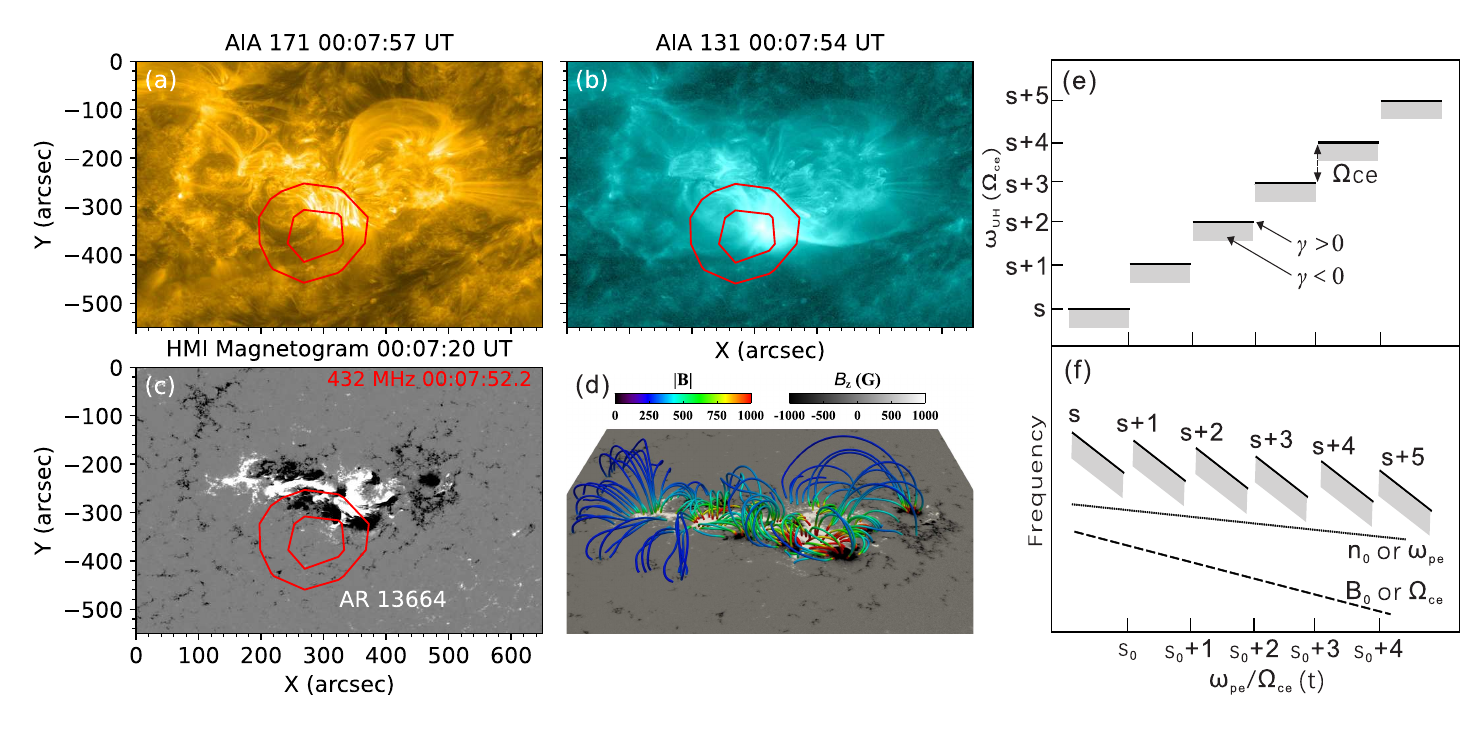}
              }
              \caption{SDO observations and magnetic field modeling of the 2024 May 9 event, along with schematic diagrams illustrating the generation of striped chains. (a) and (b) AIA 171  \AA \ and 131 \AA \ EUV images of AR 13664. (c) HMI line-of-sight magnetogram. Overlaid contours show the simultaneous DART radio sources at 432 MHz (70\% and 90\% of the maximum brightness temperature). The DART imaging time and frequency are marked in the dynamic spectrum of Figure \ref{Fig:figure3}. (d) NLFFF-extrapolated magnetic field lines. (e) Variations of the UH mode frequency as a function of $\omega_{pe}/\Omega_{ce}$ (t). Shaded regions represent absorptions, and $\gamma$ denotes the growth rate. (f) Schematic of chain formation due to variations of $\omega_{pe}/\Omega_{ce}$ (t).
              }
   \label{Fig:figure7}
   \end{figure*}

 \begin{figure*}
   \centerline{\includegraphics[width=0.4\textwidth,angle=90]{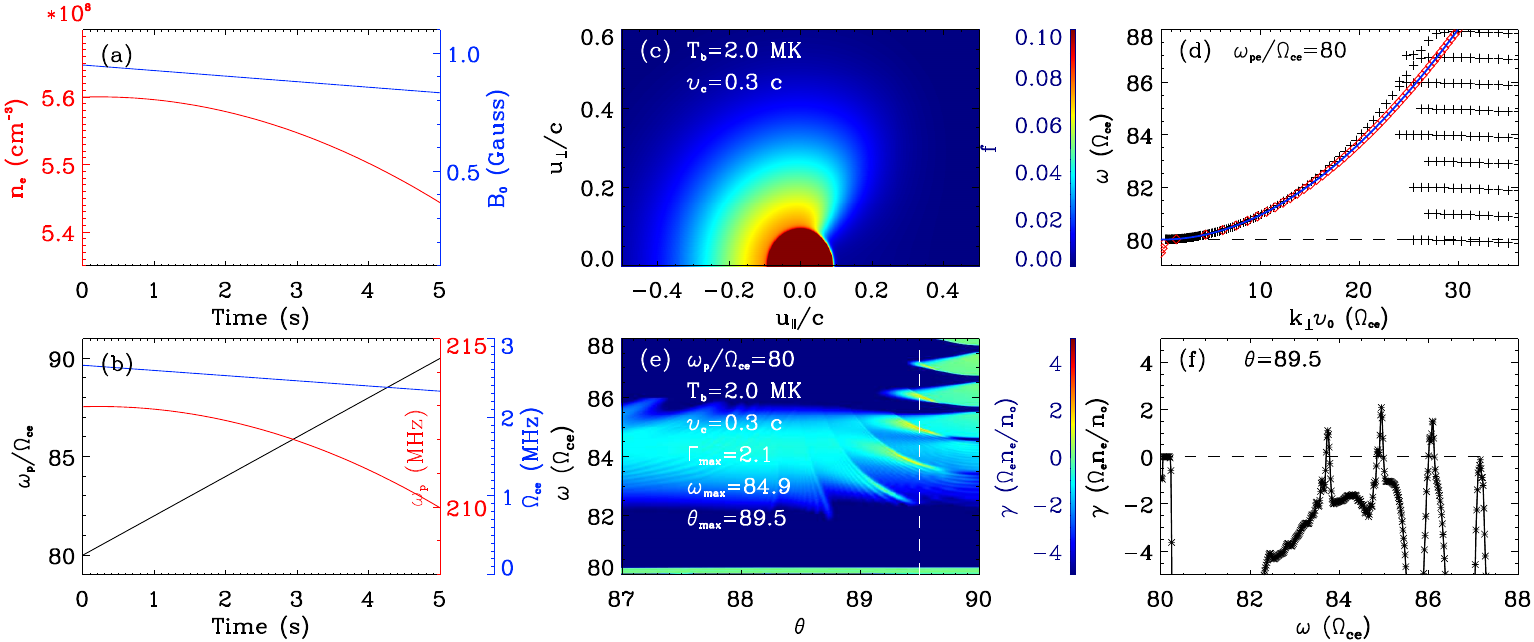}
              }
              \caption{Model setup and Z-mode (UH) wave analysis. (a) Temporal evolution of the plasma density $n_0$ (red) and magnetic field strength $B_0$ (blue) in the radio source region. (b) Profiles of plasma frequency $\omega_{pe}$, electron cyclotron frequency $\Omega_{ce}$, and their ratio $\omega_{pe}/\Omega_{ce}$ (red, blue, and black lines, respectively). (c) Electron distribution adopted for the linear growth rate calculation: background Maxwellian ($T_0$ = 2 MK) and energetic electrons with a loss-cone distribution (average velocity $v_e = 0.3c$). (d) Dispersion relations of Z-mode waves in warm plasmas with $\omega_{pe}/\Omega_{ce}$ = 80 for perpendicular propagation, calculated using fluid approximation (red circles) and kinetic theory (black plus signs). The blue line corresponds to the dispersion curve of UH mode, $\omega^2=\omega^2_p+\Omega^2_{ce}+3k^2_\perp v^2_0$, and the black dashed line indicates the UH frequency ($\omega_{UH}^2 = \omega_{pe}^2 + \Omega_{ce}^2$). (e) Growth rate distribution of Z-mode waves in ($\omega, \theta$) space; positive values indicate growth and negative values damping. (f) Growth rate of Z-mode waves at fixed propagation angle $\theta = 89.5^{\circ}$.
              }
   \label{Fig:figure8}
   \end{figure*}

 \begin{figure*}
   \centerline{\includegraphics[width=0.45\textwidth,angle=90]{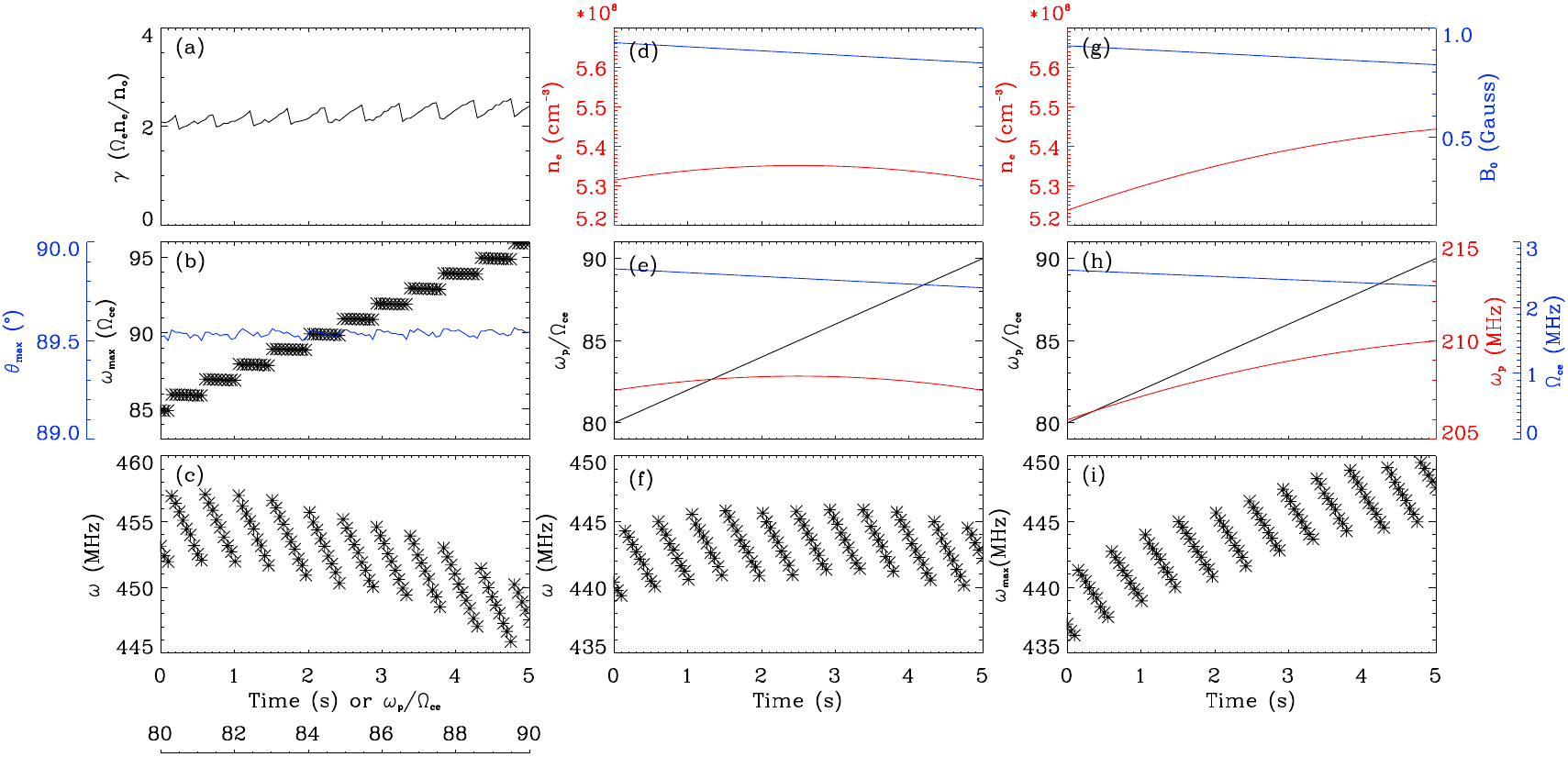}
              }
              \caption{(a) Maximum growth rate of the UH mode as a function of $\omega_{pe}/\Omega_{ce}$ (t). (b) Corresponding real frequency and propagation angle at the maximum growth rate. (c) Radiation spectra produced by mode conversion of UH waves. The middle and right columns show the resulting spectra for different temporal variations of $n_0$ and $B_0$.
              }
   \label{Fig:figure9}
   \end{figure*}

 \begin{figure*}
   \centerline{\includegraphics[width=0.85\textwidth]{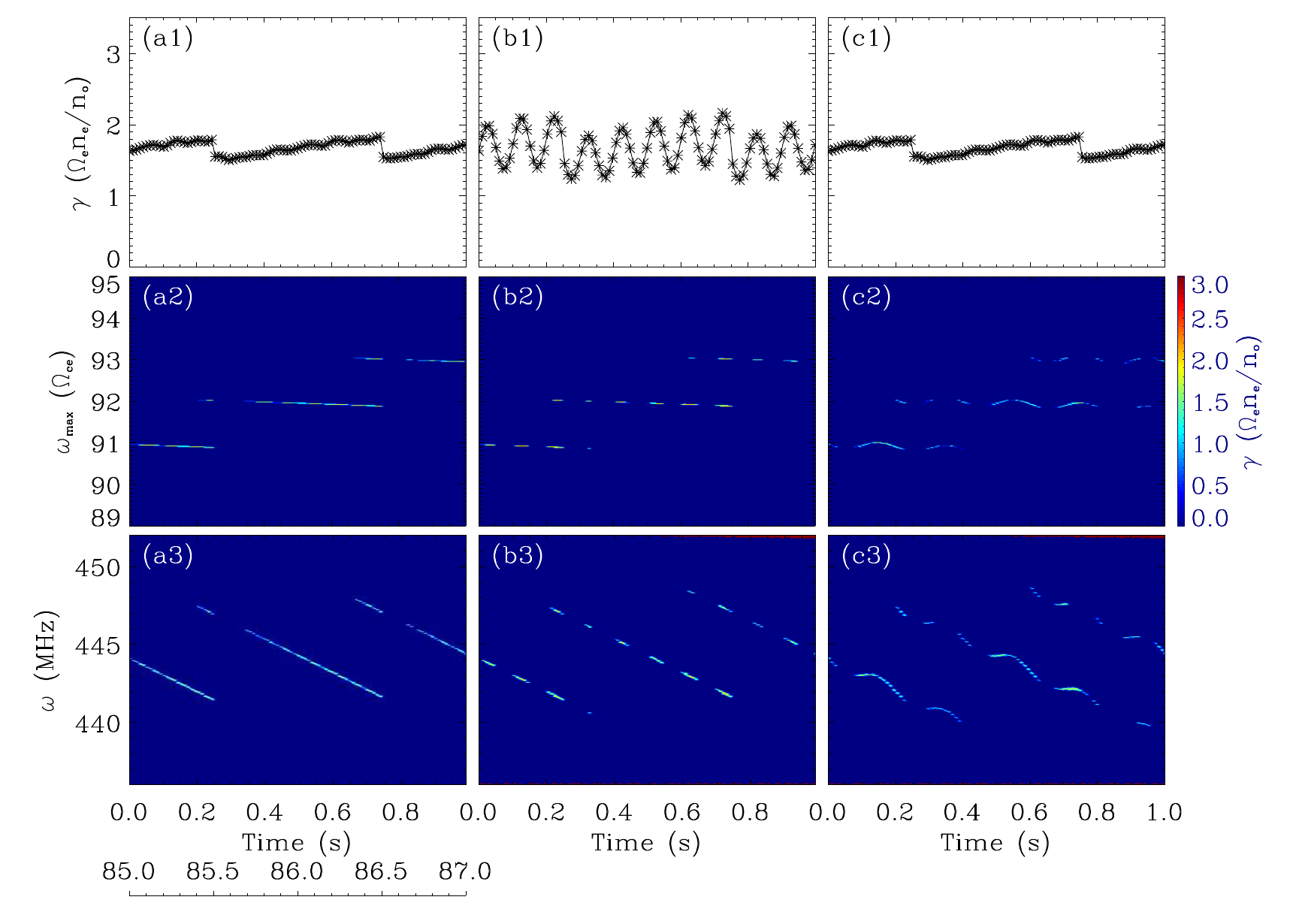}
              }
              \caption{UH mode growth rates and radiation spectra under three cases: no modulation (left column), modulation of the energetic electron density (middle column), and modulation of the magnetic field direction (right column). From top to bottom: maximum               growth rate versus $\omega_{pe}/\Omega_{ce}$ (85--87, corresponding to 0--1 s); frequency distribution; radiation spectra.
              }
   \label{Fig:figure10}
   \end{figure*}

 \begin{figure*}
   \centerline{\includegraphics[width=0.9\textwidth]{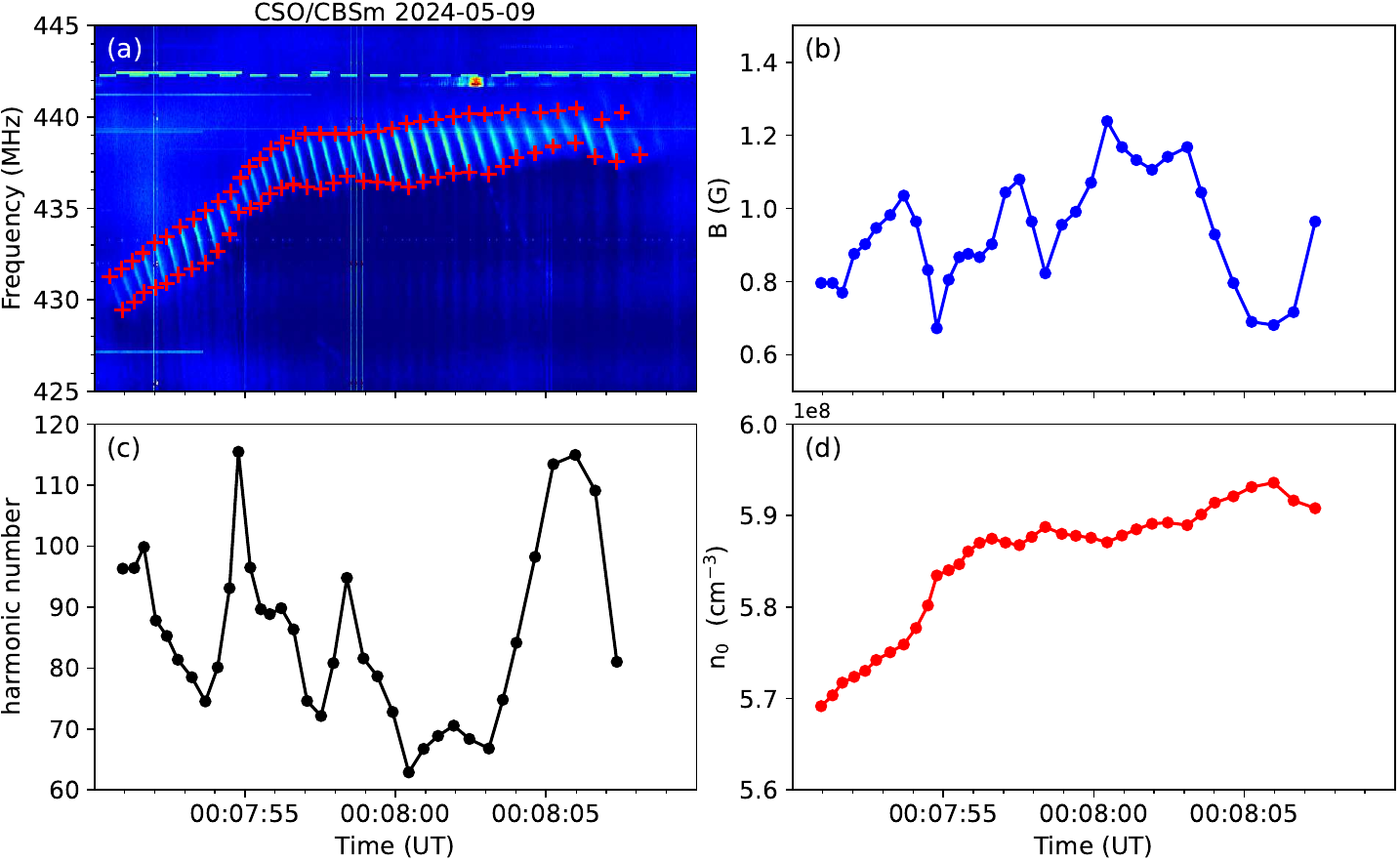}
              }
              \caption{(a) Periodic narrow-band chained stripes event on 2024 May 9 (Figure \ref{Fig:figure3}(b)), with ``$+$'' symbols mark the start and end of each stripe. (b) Magnetic field strength obtained from stripes. (c) Harmonic numbers of the excited UH waves. (d) Plasma density in the source region.
              }
   \label{Fig:figure11}
   \end{figure*}

 \begin{figure*}
   \centerline{\includegraphics[width=0.9\textwidth]{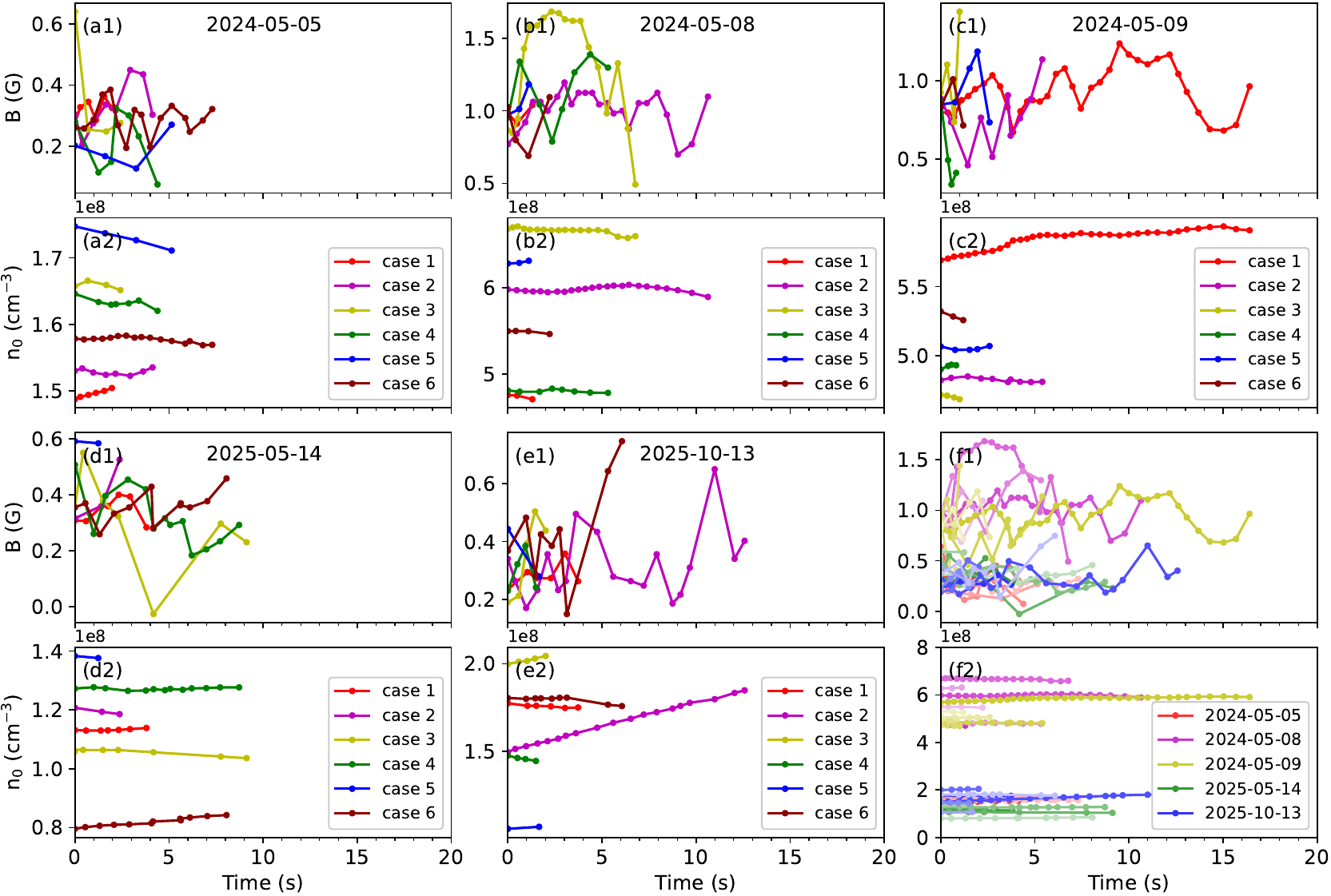}
              }
              \caption{Magnetic field strengths and plasma densities obtained from the proposed model and observations for five events, each represented by six chained stripes (Figures \ref{Fig:figure1}--\ref{Fig:figure5}, panels (d)--(i)). Panels (f1) and (f2) summarizes all thirty chains.
              }
   \label{Fig:figure12}
   \end{figure*}

\end{document}